\documentclass[12pt]{iopart}

\usepackage{graphicx}
\usepackage{dcolumn}
\usepackage{bm}
\usepackage{algorithm}
\usepackage{enumerate}
\usepackage{algpseudocode}
\usepackage{algorithmicx}
\usepackage{acronym}
\usepackage{xcolor}
\usepackage[normalem]{ulem} 
\usepackage{amsthm, amssymb, amsfonts}
\usepackage{caption}
\usepackage{subcaption}
\usepackage{hyperref}
\usepackage{cleveref}

\newcommand{\fd}{$T_{E}$}
\newcommand{\linf}{$L_{\infty}$} 
\newcommand{\ltwo}{$L_{2}$}

\newcommand{\gra}{\texttt{GR-Athena++}}
\newcommand\be{\begin{equation}}
\newcommand\ee{\end{equation}}
\newcommand\ba{\begin{eqnarray}} 
\newcommand\ea{\end{eqnarray}} 

\newcommand{\added}[1]{#1} 

\acrodef{AMR}{adaptive mesh refinement}
\acrodef{BNS}{binary neutron star}
\acrodef{BBH}{binary black hole}
\acrodef{NS}{neutron star}
\acrodef{BH}{black hole}
\acrodef{EOS}{equation of state}
\acrodef{NR}{numerical relativity}
\acrodef{AMR}{adaptive mesh refinement}
\acrodef{GW}{Gravitational Wave}
\acrodef{GR}{General Relativity}
\acrodef{MB}{mesh block}

\begin{document}

\title{Adaptive mesh refinement in binary black holes simulations}

\author{Alireza Rashti$^{1,2}$, Maitraya Bhattacharyya$^{1,2}$, David
Radice$^{1,2,3}$\footnote{Alfred P.~Sloan Fellow}, Boris Daszuta$^{4}$,
William Cook$^{4}$, and Sebastiano Bernuzzi$^{4}$}

\address{$^{1}$
		Institute for Gravitation and the Cosmos,
		The Pennsylvania State University,
		University Park, PA 16802, USA}
\address{$^{2}$
		Department of Physics,
		The Pennsylvania State University,
		University Park, PA 16802, USA}
\address{$^{3}$
		Department of Astronomy \& Astrophysics,
		The Pennsylvania State University,
		University Park, PA 16802, USA}
\address{$^{4}$
  Theoretisch-Physikalisches Institut,
  Friedrich-Schiller-Universit{\"a}t Jena, 07743, Jena, Germany}

\ead{dur566@psu.edu}

\begin{abstract}
    We discuss refinement criteria for the 
    Berger-Rigoutsos~(block-based) refinement algorithm in our 
    numerical relativity code \gra~in the context of binary black hole 
    merger simulations. We compare three different strategies: the 
    ``box-in-box'' approach, the ``sphere-in-sphere'' approach and a 
    local criterion for refinement based on the estimation of 
    truncation error of the finite difference scheme. We extract and 
    compare gravitational waveforms using the three different mesh 
    refinement methods and compare their accuracy against a calibration 
    waveform and demonstrate that the sphere-in-sphere approach 
    provides the best strategy overall when considering computational 
    cost and the waveform accuracy. Ultimately, we demonstrate the 
    capability of each mesh refinement method in accurately simulating 
    gravitational waves from binary black hole systems --- a 
    crucial aspect for their application in next-generation detectors.
    We quantify the mismatch achievable with the different strategies
    by extrapolating the gravitational wave mismatch to higher resolution.
\end{abstract}

%
\vspace{2pc}
\newline
\\
%
\submitto{\CQG}
%
%
%
\section{Introduction} \label{sec:introduction}

An essential tool in the gravitational wave astrophysicist's arsenal is 
accurate waveform models for physical systems of interest. For \ac{BBH} 
merger systems, the most accurate waveform have been obtained by 
solving the full Einstein's equations describing the spacetime dynamics 
of these compact objects~\cite{Mroue:2013xna, Boyle:2019kee, 
Jani:2016wkt, Healy:2017psd, Hamilton:2023qkv, Ferguson:2023vta}. These 
simulations are then used to inform waveform models used for 
gravitational wave data analysis, e.g.,
\cite{Taracchini:2013rva, 
Pan:2013rra, Hannam:2013oca, Khan:2015jqa, Babak:2016tgq, Bohe:2016gbl, 
Mehta:2017jpq, London:2017bcn, Khan:2018fmp, Nagar:2018zoe,Varma:2018mmi,
Nagar:2020xsk,Garcia-Quiros:2020qpx,Pratten:2020ceb,
Gamba:2021ydi,Pompili:2023tna,vandeMeent:2023ols,Ramos-Buades:2023ehm,
Thompson:2023ase,Nagar:2023zxh}.
In particular, the waveform models are needed to 
extract physical parameters of astrophysical system from their 
gravitational wave data. For this reason, accurate waveform models are 
crucial for determining the masses and spins of the black hole 
precisely with low degree of uncertainty. Moreover, these models play 
an important role in scrutinizing fundamental physics, particularly in 
the examination of general relativity within the non-linear 
regime~\cite{LIGOScientific:2016lio, LIGOScientific:2018dkp}. In the 
realm of parameter estimation, the introduction of bias is a notable 
concern stemming from the use of inaccurate numerical 
models~\cite{PhysRevResearch.2.023151}. This apprehension gains 
significance with the advancing sensitivity of the current generation 
LIGO and Virgo detectors ~\cite{LIGOScientific:2016aoc, 
LIGOScientific:2016sjg, LIGOScientific:2017bnn, LIGOScientific:2017vox, 
LIGOScientific:2017ycc, LIGOScientific:2017ync}. As we anticipate the 
era of next-generation detectors like the Laser Interferometer Space 
Antenna (LISA) ~\cite{LISA:2017pwj}, the Einstein 
Telescope~\cite{Maggiore:2019uih}, and Cosmic 
Explorer~\cite{Reitze:2019iox}, the demand for highly accurate 
theoretical waveforms using \ac{NR} becomes increasingly indispensable. 
To this end, considerable efforts have been dedicated to the 
development of \ac{NR} infrastructures such as 
\texttt{BAM}~\cite{Bruegmann:1996kz, Bruegmann:2006ulg}, 
\texttt{bamps}~\cite{Renkhoff:2023nfw}, 
\texttt{Dendro-GR}~\cite{Fernando:2022php}, \texttt{Einstein 
Toolkit}~\cite{EinsteinToolkit:2023_05, Schnetter:2003rb, 
Loffler:2011ay}, \gra~\cite{Daszuta:2021ecf}, 
\texttt{GRChombo}~\cite{Clough:2015sqa}, 
\texttt{Nmesh}~\cite{Tichy:2022hpa}, 
\texttt{SENR/NRPy+}~\cite{Ruchlin:2017com}, 
\texttt{Simflowny}~\cite{Palenzuela:2018sly}, 
\texttt{SpEC}~\cite{Kidder:2000yq}, 
\texttt{SpECTRE}~\cite{Kidder:2016hev}, among others.

In \ac{NR} codes refinement approaches such as \ac{AMR} and static mesh 
refinement play an important role for efficiency and accuracy of the 
numerical simulations. Unlike static mesh refinement, \ac{AMR} 
introduces the capability for dynamic changes in the computational 
domain resolution, enabling more efficient simulations. Typically, 
there are two \ac{AMR} paradigms: patch-based \ac{AMR} and block-based 
\ac{AMR}. In the patch-based \ac{AMR} the computational domain is 
covered with overlapping subdomains each with different resolutions. 
One of the most common implementations of this paradigm is the 
Berger-Oliger algorithm~\cite{Berger:1984zza}. This implementation 
requires a complex parallelization algorithm for meshes of multiple 
levels to communicate with each other. Moreover, quantities at 
different levels of refinement can possess different values, thereby 
leading to different dynamics in the same region~\cite{Stone2020TheAA}. 
This can be problematic if the evolution is terminated due to 
catastrophic accumulation of error on the coarse grids, where the 
dynamics is severely under resolved. An alternative to this paradigm is 
block-based \ac{AMR}~\cite{Stout1997AdaptiveBA} where subdomains of 
different levels do not have regions of overlap except at shared 
boundaries. In \gra, we use a Berger-Rigoutsos~\cite{120081} algorithm 
for mesh refinement, a detailed exposition of which is provided 
in~\cite{Stone2020TheAA, Daszuta:2021ecf}.

In this work, we focus our attention on \ac{BBH} mergers using three 
different criteria in the context of block-based \ac{AMR} in our code. 
For the first type, we emulate the popular ``box-in-box" approach with 
a hierarchy of \acp{MB} of different spatial resolutions surrounding 
the black hole punctures and moving with it.  This requires tracking of 
the position of the black hole punctures throughout the simulation. The 
criteria for refining or coarsening a part of the numerical grid is 
controlled by its distance from the puncture locations, with the 
maximum norm~\linf~being chosen to compute distances. The second 
criterion emulates a ``sphere-in-sphere" approach in which the norm is 
replaced by the Euclidean norm \ltwo. 
The third criterion is chosen based 
on an estimation of the local truncation error of the finite difference 
scheme. This approach does not require knowledge of the positions of 
the black holes and can be generalized to tackle other problems. We 
remark that Ref.~\cite{Radia:2021smk} contrasted the cell-centered AMR 
criteria of \texttt{GRChombo} and the box-in-box approach of 
\texttt{Lean}. Our study compares box-in-box and truncation-error 
driven AMR in the context of the vertex-centered \texttt{GR-Athena++} 
code. We find that by using \ltwo~and \linf~criteria, it might be 
possible produce waveforms with mismatch of~$\mathcal{O}(10^{-7})$ 
thereby making these promising approaches for producing highly accurate 
\ac{NR} waveforms for use with next-generation detectors.

The paper is structured as follows:
In \Cref{sec:amr}, we provide a concise discussion on the grid setup 
and the refinement algorithm employed in~\gra. 
Moving on to \Cref{sec:indicator}, we delve into the indicators within our code 
that trigger refinement, including the norm-based refinement criteria, namely,
\linf~and \ltwo~methods as well as the local approximation of the finite difference 
truncation error, so-called the \fd~method.
The objective of \Cref{sec:results} is to scrutinize the efficiency and accuracy 
of each \ac{AMR} criterion.
Initially, in \Cref{sec:efficiency}, we conduct a comparative analysis of the 
efficiency of each \ac{AMR} criterion and address the challenges posed 
by the \fd~method.
Subsequently, in \Cref{sec:convergence}, we explore the convergence tests of 
the waveforms generated using the three refinement methods.
Following this, \Cref{sec:waccuracy} delves into the discussion of 
their accuracy where it assess the potential of employing these refinement methods 
for high-fidelity \acp{GW}.
The paper concludes in \Cref{sec:discussion} with remarks on future directions
for research.
We use geometrized units~$ G = c = 1$, where $G$ is the constant of 
gravity, and $c$ is the speed of light, throughout the paper.

\section{Adaptive mesh refinement in \gra}
\label{sec:amr}
\subsection{Grid setup}
\label{sec:grid_setup}

In order to increase the computational efficiency through parallelization,
the computational grid~(numerical domain) in~\gra~is decomposed into a number of subdomains
which are referred to as~\acfp{MB}.
Furthermore, the arrangement of~\acp{MB} is internally represented as a tree data 
structure: a binary tree for one-dimensional problems, a quadtree 
in two dimensions and an octree in three dimensions~\cite{Stone2020TheAA,Daszuta:2021ecf}.
This representation allows us to represent relationships between ``parent"
and ``child"~\ac{MB} conveniently when mesh refinement is used.

The resolution on the root grid is controlled by specifying the number 
of grid points in each direction. The base grid is then subdivided into 
\acp{MB}, each having a predetermined number of grid points, which 
should be a divisor for the number of points in the root grid. The 
\acp{MB} can then be subdivided recursively to create the full-\ac{AMR} 
grid starting from the root grid. In particular, for a~$d$ dimensional 
problem, each~\textit{refinement} operation increases the level of 
refinement by 1 and subdivides an~\ac{MB} into~$2^d$ ``children" 
subject to the fulfillment of a refinement criterion. The converse is 
achieved during coarsening~(``de-refinement"), which depends on a 
coarsening criterion. We refer the reader to a detailed description 
\added{and illustration}
of the tree data structure, Z-ordering, and communication 
to~\cite{Stone2020TheAA, Daszuta:2021ecf}.

\subsection{Refinement algorithm}
\label{sec:refinement_algo}

At every time step, each \ac{MB} is tagged for refinement, coarsening
or for keeping it as is, subject to a refinement criterion. This typically involves
computing a scalar number for each \ac{MB}, which is then compared against
a given range provided by the user. The \acp{MB} are then refined or
coarsened subject to refinement constraints which respect the fact that
each~\ac{MB} can have neighbors whose refinement level are at most 
$\pm 1$ from itself~\cite{Stone2020TheAA}. In the case of refinement, an
\ac{MB}~(in three dimensions) is subdivided into 8 ``children" \acp{MB}
with edges of half length as its ``parent". A prolongation operation 
interpolates data from the ``parent" to the ``children". Similarly,
sibling grids which are all tagged for coarsening are consolidated
into a single \ac{MB}, with data being copied to a coarser grid by a 
restriction operation.
\added{\cite{Stone2020TheAA, Daszuta:2021ecf} provide additional
description and illustration for refinement algorithm in \gra.}

\subsection{Prolongation and restriction}
\label{sec:prolongationrestriction}

Boundary conditions at each \ac{MB} are implemented in~\gra~by the use 
of ghost zones. Each~\ac{MB} can communicate with its neighbors by 
exchanging data between their overlapping ghost zones. For 
neighboring~\acp{MB} which are not at the same refinement level, 
prolongation is performed after the data is exchanged (for coarse to 
fine communication), while restriction is performed prior to data 
exchange (for fine to coarse communication). Similarly, when an \ac{MB} 
undergoes refinement or coarsening, these same operations are performed 
to store data on the resulting \acp{MB}.

To map data from a coarser~\ac{MB} to a finer~\ac{MB}, a prolongation 
operation is performed. For a vertex centered variable, this is done 
using Lagrange polynomial interpolation. A smooth function~$f$, sampled 
over an equispaced grid of $2 N + 1$ points in an \ac{MB} can be 
uniquely written as
\ba
	\tilde{f}(x) = \sum_{i= - N}^{N} L_{i} (x) f_i,
\ea
where~$L_{i}(x)$ are the Lagrange cardinal polynomials 
satisfying~$L_i(x_j) = \delta_{ij}$ and~$f_i$ are the values of~$f$ 
at~$x_i$. This interpolant can then be used to compute $f$ on the finer 
mesh. Similarly, to map data from a finer~\ac{MB} to a coarser one, a 
restriction operation is performed. This is trivial in our case, 
because with 2:1 grid refinement fraction and vertex-centered sampling 
of the data, the coarser grid points form a subset of the finer grid, 
and a direct copy operation can be employed.

\section{Refinement indicators} \label{sec:indicator}

\subsection{Tracker-based AMR} \label{sec:amr_linf_l2}

In our simulations, the position of each black hole at every time
step can be determined approximately by tracking the location
of the punctures~\cite{Campanelli:2005dd}. This is done by
solving an ordinary differential equation in addition to the Z4c
equations~\cite{Daszuta:2021ecf}:
\ba
\frac{d x^i_p}{dt} = - \beta^i|_{\bm{x}_p} (t),
\ea
where $\beta^i|_{\bm{x}_p}$ is the shift vector evaluated at the 
position of the puncture ~$\bm{x}_p$. Knowledge of puncture locations 
allows us to refine in the region near the black holes, where the 
spatial derivatives of gravitational fields are changing rapidly in 
this regime of strong gravity. Following the well known box-in-box 
approach used in many numerical relativity 
codes~\cite{Bruegmann:2006ulg,Cao:2008wn,Loffler:2011ay,Clough:2015sqa}, 
we design criteria where the punctures are covered by a non-overlapping 
arrangement of \acp{MB} with the increasing refinement levels closer to 
the location of the punctures.

The criterion for refinement in this case is not local in nature, since 
\acp{MB} are taggered for refinement/derefinement not only on the basis 
of the value of the fields in their interior, but also on the basis of 
the puncture positions at every time step. Given the coordinate 
position of the~$p$-th puncture~$x^i_p$, the desired physical 
refinement level of an \ac{MB} is calculated by computing the minimum 
distance between the location of the puncture and the location of the 
\ac{MB}. This is determined by calculating the minimum distance between 
the puncture and the vertices of a rectangular block centered at the 
\ac{MB} center, where the rectangular block's edges are set at a length 
of $1/4$ of the size of the edges of the \ac{MB}. 
\added{The purpose of this specific rectangular block at the center, 
with its edge length, is to imitate the box-in-box structure of
of the BAM code~\cite{Daszuta:2021ecf} when using the \linf{}norm.}
The minimum distance, 
considering all puncture locations is stored in $d$:
\ba
	d = \textrm{min} || \bm{x}_p -  \bm{x}_{v} ||_k,
\ea
where the minimum is considered for all punctures with locations 
$\bm{x}_p$ between all vertices $\bm{x}_v$ of the rectangular block. 
Here~$k$ denotes the norm used to compute distances such as Euclidean 
norm \ltwo~or the maximum norm \linf. The result is compared with the 
current refinement level of the \ac{MB}, after which the algorithm 
decides to refine, coarsen or keep as is, subject to the constraint 
that the simulation has a maximum refinement level~$l_{max}$.
\added{Additionally, \acp{MB} with $d$ greater than half of
the computational grid extent $S$ are coarsen.
In other words, refinement of \acp{MB} that reside diagonally with respect
to the puncture position at the edge of the computational grid are
discouraged.}
The detailed criterion is presented in \Cref{alg:l2linf}.

\begin{figure}[h]
 \begin{algorithm}[H]
 \caption{Static \linf/\ltwo-norm criterion:
  \\
  \emph{input:} \added{half of the computational grid extent} $S$, $l_{\max}$, and a root grid.}
 \begin{algorithmic}[1]
 \State Start with the root grid;
 \For{ each \ac{MB} with center position $\bm{x}_{\rm mb}$}
 \State Find $\bm{x}_{v_j}$, $j = 1,\ldots,8$, the vertices of the
 rectangular block centered at $\bm{x}_{\rm mb}$;
 \For{ each puncture with position $\bm{x}_p$}
 \State Find $d_p = \min || \bm{x}_{p}-\bm{x}_{v_j}||_k$ for $j=1,\ldots,8$;
 \EndFor
 \State d $\leftarrow \min_{p} d_{p}$;

 \If {$\frac{S}{d} < 1$}
 	\State Coarsen;
 \ElsIf {$\lfloor \log_2(\frac{S}{d})\rfloor < l_{\max}$}
 	\State Refine;
 \ElsIf {$\lfloor \log_2(\frac{S}{d})\rfloor > l_{\max}$ $\rightarrow$}
 	\State Coarsen;
 \Else
 	\State Keep it as is;
 \EndIf
 \EndFor
 \end{algorithmic}\label{alg:l2linf}
 \end{algorithm}
\end{figure}

\begin{figure}
	\centering
	\begin{subfigure}{.5\textwidth}
		\centering
		\includegraphics[width=\linewidth]{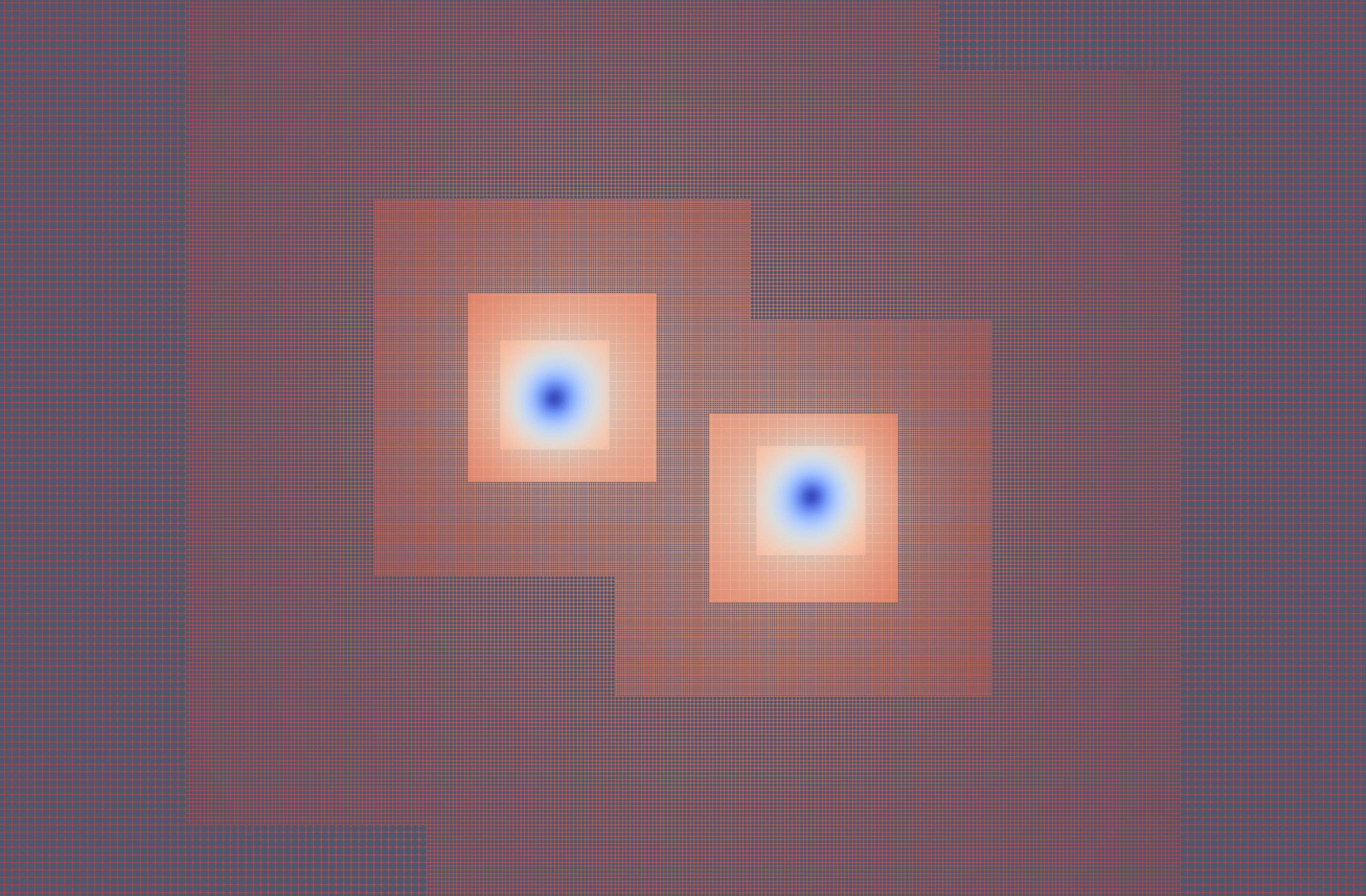}
	\end{subfigure}%
	\hspace*{1mm}
	\begin{subfigure}{.5\textwidth}
		\centering
		\includegraphics[width=\linewidth]{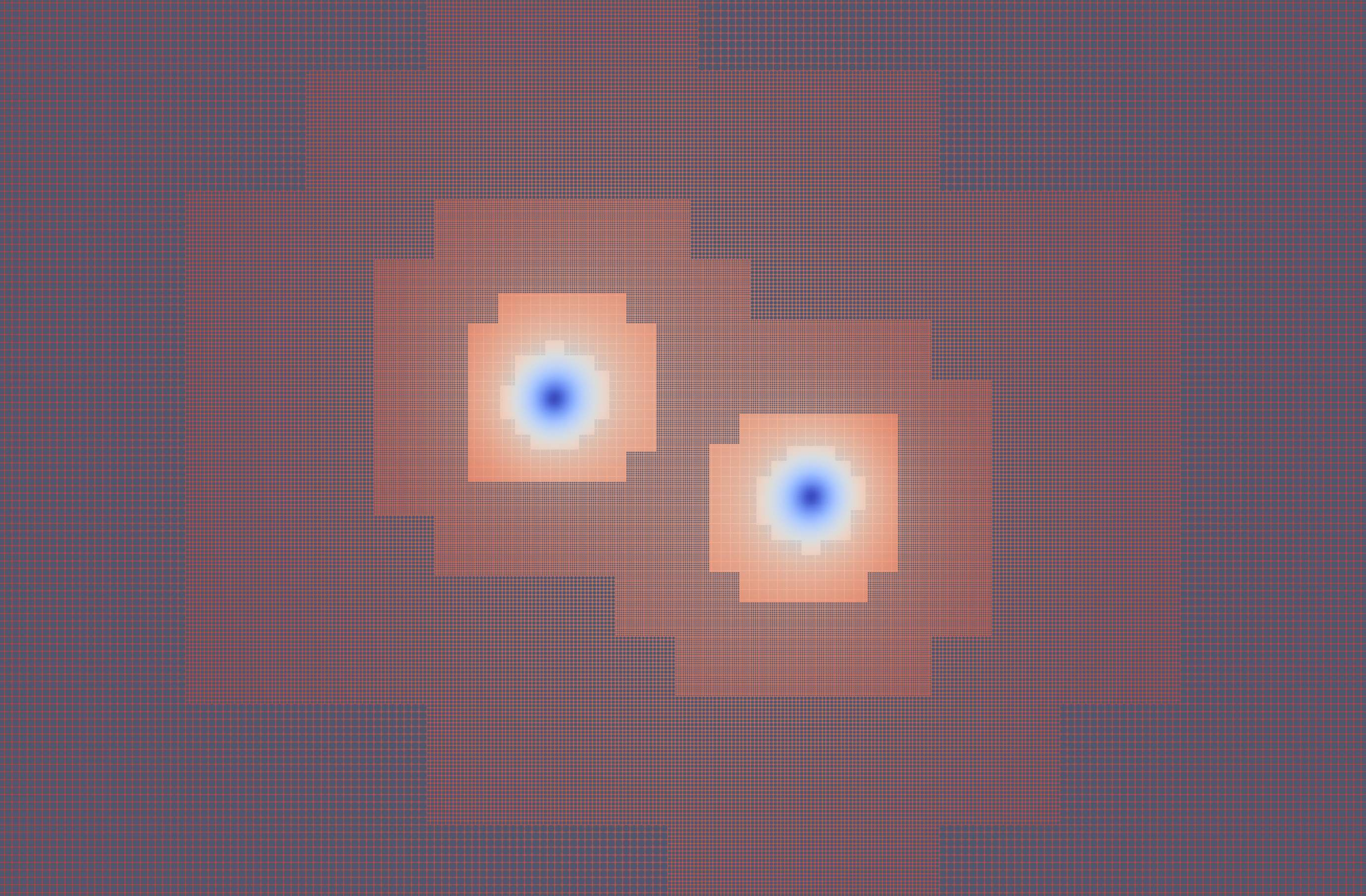}
	\end{subfigure}%
	\caption{%
  The grid structure in the $x-y$ plane for an equal mass 
  \ac{BBH} simulation. The snapshots are taken at the same evolution time. 
	Distinct refinement levels are illustrated 
	using varying colors, with the lightest shade indicating the finest level. 
  Dark blue spots represent the punctures and their immediate surroundings.
  \emph{Left panel:} Deploying \linf-norm based refinement.
	The arrangement of \acp{MB} reveals a distinctive box-in-box pattern.
  \emph{Right panel:} Using \ltwo-norm based refinement. 
	A discernible sphere-in-sphere pattern is seen.
  }
	\label{fig:amr_linfl2}
\end{figure}

The choice of norm determines the exact arrangement of \acp{MB} around the
puncture. In our simulations we consider the \linf~and \ltwo~norms. As seen
in the left and right panels of \Cref{fig:amr_linfl2}, using~\linf~leads to
a ``box-in-box" arrangement of \acp{MB} around each puncture while
choosing~\ltwo~leads to a ``sphere-in-sphere" arrangement. 
From a geometric perspective, when considering the same distance relative to the 
puncture, the \linf~method delineates an imaginary box and proceeds to label 
all the \acp{MB} located within this box for the coarsening. 
This results in a ``box-in-box" configuration. 
In contrast, the \ltwo~method defines a sphere and tags all the contained
\acp{MB} within it, resulting in a ``sphere-in-sphere" arrangement.
Furthermore, it is worth noting that the imaginary sphere created by the
\ltwo~method is actually inscribed within the box produced by the \linf~method. 
As we will explore in \Cref{sec:efficiency}, this characteristic renders 
the \ltwo~method more efficient compared to the \linf~method.

%

\subsection{Error-based AMR} \label{sec:amr_fd}

As an alternative criterion to the tracker-based \ac{AMR}, we now
consider a method that is based on the truncation error of the finite
difference derivative scheme and we call it the~\fd~method. This method,
as opposed to the methods of \Cref{sec:amr_linf_l2}, i.e., \linf~and~\ltwo,
does not require the information about puncture position at each time step
and hence, advantageously, is a local operation for each \ac{MB}.

In this work, we use a sixth-order accurate finite different method for the spatial
derivative setup in \gra. Hence, one can estimate the local truncation error of the
derivatives analytically.
Consider representation of
a $C^{n+1}$ continuous function $f$ over ${x_0,x_1,...,x_n}$ distinct grid
points from the interval~$\mathcal{I}$~\cite{burden2011numerical}
\ba
  f(x) = \sum_{i=0}^{n} f(x_i) L_i(x) +
    \frac{(x-x_0)\cdots(x-x_n)}{(n+1)!}f^{(n+1)}(\xi),
  \label{eq:lagrange_fit}
\ea
where $L_i(x)$ are the Lagrange coefficient polynomials and $\xi$ is some
value in the interval $\mathcal{I}$.  The first derivative of~$f$ at $x_j$
is given by
\ba
  f^{\prime}(x_j) = \sum_{i=0}^{n} f(x_i) L_i^{\prime}(x_j) +
    \frac{f^{(n+1)}(\xi(x_j))}{(n+1)!} \prod_{i=0, i\neq j}^{n} (x_j - x_i).
  \label{eq:deriv_lagrange_fit}
\ea
As such in \gra~one can model the spatial derivative error
budget as $C |f^{(7)}(\xi)| {h}^6 $,
for some constant $C$ and the grid space $h$ at each \ac{MB}.
Accordingly,
we compute $\left|f^{(7)}(\xi) {h}^6\right|$ at each \ac{MB} point and
consider its maximum bound as a proxy for the truncation error $\mathcal{E}$
in the given \ac{MB}, i.e.,
\ba \label{eq:truncation.error}
\mathcal{E} = \max\left|f^{(7)}(\xi) {h}^6\right|.
\ea
The field we use to calculate the error $\mathcal{E}$ is the conformal
factor~$\chi$ of the Z4c system~\cite{Bernuzzi:2009ex,Hilditch:2012fp}. 
This choice is motivated by the fact that in the single puncture 
\added{initial data}
solution of
Einstein's field equations, with the bare mass~$m_0$, 
the conformal factor $\chi$ is~\cite{alcubierre_book}
\ba
  \chi = 1+\frac{m_0}{2 r},
\ea
and hence, $\chi^{(7)} \rightarrow \infty$ as $r \rightarrow 0$ as well 
as $\chi^{(7)} \rightarrow 0$ as $r \rightarrow \infty$, where 
$\chi^{(7)}$ is the seventh-order derivative of $\chi$ with respect to 
$r$. 
\added{It is worth noting that during dynamical evolution when the moving puncture 
gauges settle down, $\chi$ at the puncture behaves as $1/\sqrt{r}$%
~\cite{Brown:2007tb} and still 
$\chi^{(7)} \rightarrow \infty$ as $r \rightarrow 0$.}
\added{Additionally, we} note that the Eq.~(\ref{eq:truncation.error}) 
is not formally 
valid at the location of the puncture, since the data is not smooth 
there. However, this choice ensures that~$\mathcal{E}$ is large close 
to the puncture and small in the weak field regime, thereby creating a 
greater concentration of refinements in the strong field regime.

Having computed the error $\mathcal{E}$, we need to set a bound to trigger
refinement or coarsening during the dynamical evolution.
This bound is a free parameter at our disposal. 
Hence, one can choose different values for more refined grid or less one.
In this work, we choose this bound by demanding that
the finite difference truncation error to be smaller than or equal to the largest
asymptotic error in our numerical schemes. 
In particular, we consider the asymptotically accumulated error of
the fourth-order Runge-Kutta method as the largest error in our scheme and
hence, we ensure that~$\mathcal{E}$ is not larger than
$O\big({h}^4\big)$. Consequently, the upper and lower bounds are 
chosen as $[a\, {h}^4,b\, {h}^4] $,
where~$a$ and $b$ are defined by the user. In practice, we chose
$a = 1$ and~$b = 10$ for all our simulations.
A summary of this process is represented in \Cref{algo:amr_fd}.

\begin{figure}[h]
 \begin{algorithm}[H]
 \caption{%
	  Truncation error criterion~(6-th order finite difference):
 \\
	  \emph{input:} the error bounds a and b in $[a,b]\, h^4$ and a root grid.}
 \begin{algorithmic}[1]
   \State Start with a root grid;
   \For{ each \ac{MB} }
     \State Compute: %
	$ \mathcal{E} = \left|\left| h^6 \; \frac{\partial^7 \chi }{\partial x^7}\right|\right|_\infty$;
     \If { $\mathcal{E} \ge b\, h^4$ }
		 	\State Refine;
		 \ElsIf{ $\mathcal{E} \le a\, h^4$ }
			\State Coarsen;
     \Else
			\State Keep as is;
     \EndIf
   \EndFor
 \end{algorithmic} \label{algo:amr_fd}
 \end{algorithm}
\end{figure}

\begin{figure}
\centering
\includegraphics[width=0.6\linewidth,clip=true]{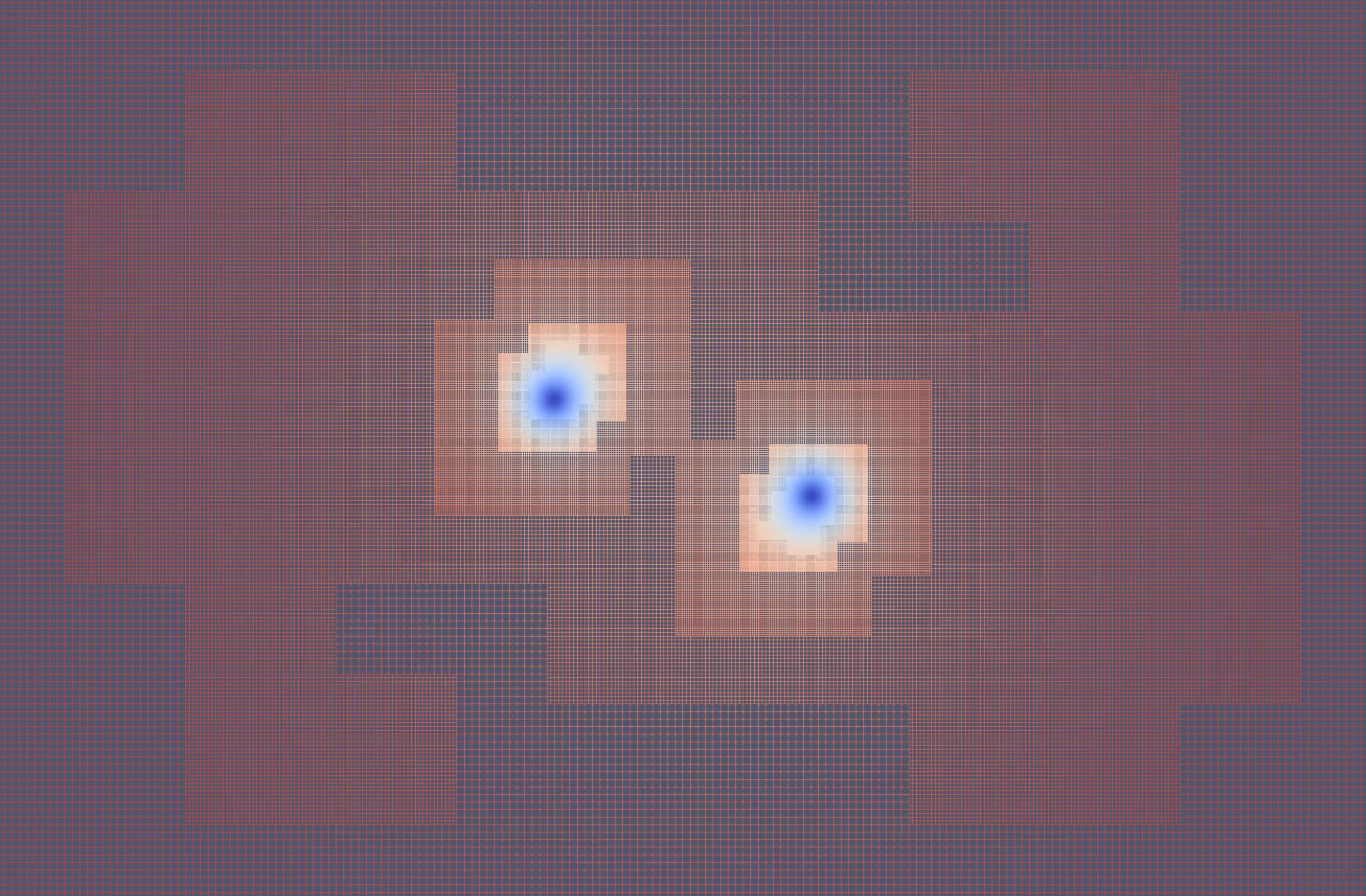}
\caption{%
	An orbital-plane snapshot of the grid structure for an equal mass \ac{BBH}
  simulation. Refinement levels are illustrated 
  using varying colors, with the lightest shade indicating the finest level. 
  Dark blue spots represent the punctures and their neighbors.
  Here the \fd~method is used for the refinement criterion.
  Comparing with \Cref{fig:amr_linfl2}, which shows the grid structure 
  at the same evolution time, we see \fd~method tends to create fewer \acp{MB}
  in the space between \ac{BBH}.
	}
\label{fig:amr_fd}
\end{figure}

Fig.~\ref{fig:amr_fd} shows a grid that is generated using the finite difference
truncation error-based refinement criterion. The criterion, as desired,
refined more near the black hole punctures. Comparing with
\Cref{fig:amr_linfl2}, we see that this criterion is
parsimonious around the puncture locations and 
the space between them. The total number of \acp{MB} created by
this method is generally less than \linf~and \ltwo~methods.
%

\section{Results}
\label{sec:results}

In this section, we consider the usefulness of our \ac{AMR} criteria, 
by scrutinizing: computational cost~(or simply the efficiency of 
criteria), the convergence order of the extracted \acp{GW}, and the 
capability of the code to produce high quality \acp{GW} for the 
next-generation detectors.

For our \ac{BBH} dynamical evolution tests the initial data is produced by 
the \texttt{TwoPunctures} code~\cite{Ansorg:2004ds}. We choose the mass 
ratio $q$ of the system $q=1$ and the total mass (at infinite 
separation) $M$ of non-spinning \acp{BH} $M = m_{1}+m_{2} = 1$. For low 
eccentricity orbits, following~\cite{Ramos-Buades:2018azo}, we take the 
separation $d$ between the binary $d=12M$ and choose the values of the 
momenta for each \ac{BH} as $|P_x| = 4.681 \times 10^{-4}/M$, $|P_y|= 
8.507 \times 10^{-2}/M$, and $|P_z| = 0.0$.

\subsection{Efficiency}
\label{sec:efficiency}

As mentioned earlier in \Cref{sec:grid_setup},~\gra~starts from a root 
grid, the configuration of which is set in the parameter file, and then 
the refinement routines begin to check the~\acp{MB} against the 
specified criterion and hence refine them and the subsequent children 
until the grid resolution satisfies the refinement criterion or exceeds 
the maximum refinement level. Note that initial data is recomputed on 
each refinement level, if refinement occurs at the beginning of the 
simulation.

Since the \linf~and \ltwo~methods are based on puncture positions and 
the initial positions are known, for a given root grid and the maximum 
refinement level, these methods can readily refine the grid structure 
in the vicinity of the \acp{BH} for an accurate interpolation of 
initial data onto the refined grid.
On the contrary, the \fd~method is not informed of the puncture 
positions a priori. When the initial data are interpolated onto the 
root grid they lack the sharp features needed to guide the \fd{} 
refinement criterion. 
Furthermore, if we start with a root grid that is properly refined 
around punctures, for instance, deploying a static mesh refinement, and 
then immediately switch on the \fd{} method after interpolating initial 
data on the grid, we observe that the \fd{} method creates excessive 
refinement,
\added{with respect to \linf~and \ltwo},
until the fields relax and junk radiation leaves the inner 
region of the computational domain, resulting in a performance drop.
To avoid these issues we use the \ltwo~refinement criterion until the 
fields are relaxed, before switching to the \fd{} method.

Another caveat we need to consider when using \fd~method is prevention 
of a grid with a too-low numerical resolution around the \ac{GW} 
extraction regions. This case can readily happen if we choose 
inappropriately the range of the controlling parameters in the \fd~method; 
namely, large values of $a$ and $b$ in $[a,b]\, h^4$ result in 
aggressive coarsening at extraction distances.
Furthermore, when using \fd{} method an excessive refinements can take 
place at the outer boundary of the computational grid. This occurs due 
to the finite domain used for our simulations with Sommerfeld boundary 
conditions imposed \cite{Hilditch:2012fp}. Consequently, some sharp and 
non-physical features can be created nearby these boundaries that 
trigger undesirable refinements in these areas.
To avoid under- or over-resolving regions in the wavezone and near the 
outer boundaries, we restrict the regions where \fd{} method is applied 
to be close to the punctures. In particular, we set the effective 
region of \fd{} method to be a solid sphere centered at the \ac{BBH} 
center of mass~(coordinate system origin) whose radius is chosen as 
large as twice the initial coordinate separation between the \acp{BH}. 
The other regions, where \fd{} method is not enabled, are only slightly 
altered as result of the early evolution with the \ltwo{} method and to 
satisfy the constraint of maximum 2:1 ratio between neighboring 
\acp{MB}.
\added{As a consequence, there is sufficient resolution at the \ac{GW} extraction
regions.}
We remark that \texttt{GRChombo} avoids these issues by 
combining multiple refinement criteria \cite{Radia:2021smk}. We do not 
investigate these options here.

\begin{figure}
  \centering
	\begin{subfigure}{.5\textwidth}
	\centering
  \includegraphics[width=0.95\linewidth,clip=true]{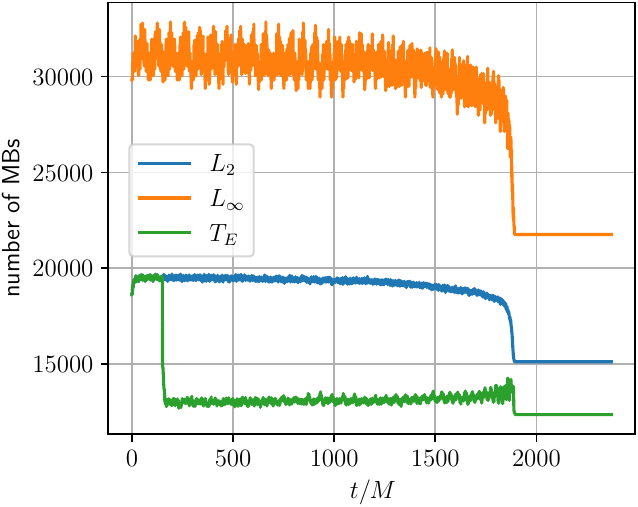}
	\end{subfigure}%
    \hfill
	\begin{subfigure}{.5\textwidth}
	\centering
  \includegraphics[width=0.95\linewidth,clip=true]{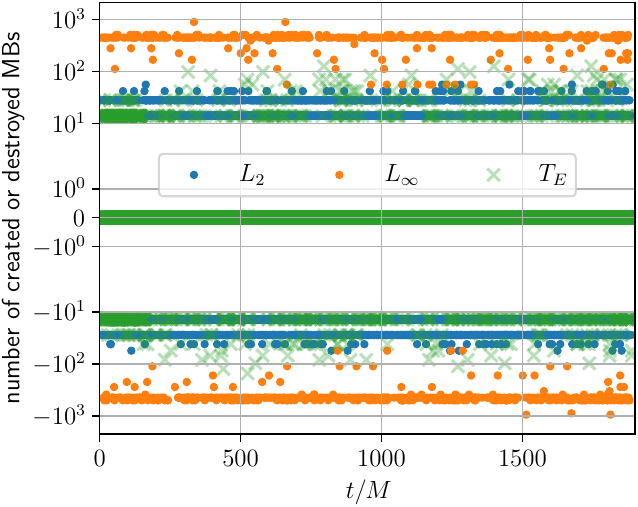}
	\end{subfigure}%
  \caption{\emph{Left panel:} total number of~\acp{MB} covering the computational
  grid at every fifth time step of the simulation.
	Different~\ac{AMR} criteria are specified with different colors. The
  largest number of~\acp{MB} belongs to the \linf~method~(orange), 
	followed by \ltwo~(blue), and then \fd~method~(green). 
	We use the \ltwo~criterion as the \fd's bootstrapping
	until $t/M\approx 150$. \emph{Right panel:} number of
  created~(positive) or destroyed~(negative)~\acp{MB} at every fiftieth time step
  for different \ac{AMR} strategies.
	\linf~strategy has the largest amplitude, the next largest is \fd~strategy 
	followed by \ltwo~strategy.
	Large oscillations of the~\ac{MB} number demand more
  load balancing and hence more computational costs.
}
\label{fig:mbs_vs_time}
\end{figure}

Fig.~\ref{fig:mbs_vs_time} shows the number of \acp{MB} at every fifth
time step of the evolution time for different \ac{AMR} criteria.
The number of points on the root grid is $256$ in each direction, \Cref{tab:amr_runs}.
We observe that even though both \linf~and \ltwo~methods are started from the same 
root grid, the \linf~method led to a larger number of \acp{MB} from the very
beginning. As mentioned in \Cref{sec:amr_linf_l2},
the \ltwo~method labels all \acp{MB} that are inside a spherical region 
for a  refinement; since this spherical region is inside the region specified by 
the \linf~method, \ltwo~method results in a fewer number of \acp{MB}.
In \fd~method we used \ltwo~method as the initial bootstrapping approach until
$t \approx 150 M$. We set this time such that the damped tail of the 
junk radiation, which arises from the choice of conformally flat initial
data~\cite{alcubierre_book}, and hence the peak amplitude of the junk radiation,
has passed from the boundaries of the domain where \fd~method applied.
We see after switching to \fd~method, that the arrangement of the \acp{MB} changes:
many of the \acp{MB} are destroyed.
Finally, we see at $t\approx 1850 M$, when the \ac{BBH} system merges
and hence there is only one \ac{BH} for \ac{AMR} criteria to work with, 
a drastic reduction in the number of \acp{MB} for all methods.
Overall, the lowest number of \acp{MB} is created by \fd~strategy followed by
the \ltwo~strategy; while the \linf~approach results in the largest number
of \acp{MB}. We observed the same behavior consistently for other resolutions as well.

Apart from the total number of \acp{MB}, an important aspect for 
performance and load balancing is the number of created or destroyed 
\acp{MB} at each time step. In the right panel of 
\Cref{fig:mbs_vs_time} we show the oscillations in the creation and 
destruction of the \acp{MB} for every fiftieth time step. In 
\Cref{fig:mbs_vs_time} at each instance of the time, a positive number 
of \ac{MB} means the number of created \acp{MB}; similarly a negative 
numbers means destroyed \ac{MB} at that time. In comparison with the 
\linf~method and \fd ~method, we see the oscillations are smallest for 
\ltwo~method and thereby there are fewer calls of the restriction and 
prolongation operators as well as the load balancing routines.

\subsection{Convergence Test}
\label{sec:convergence}

%
\begin{table}
\caption{%
Grid setups considered in this study. The computational grid covers a 
region $[-2576\, M,\ 2576\, M]^{3}$. We use $12$ refinement levels. 
Each \ac{MB} contains $16^{3}$ grid points. The total simulation time 
is $t \approx 2370\, M$. In \gra, mesh resolution specifies the number 
of grid points on the root grid in each direction. The grid space, $h$, 
indicates the resolutions at the finest level.}
\begin{center}
\begin{tabular}{lll}
\hline
\ac{AMR} criterion & mesh resolution & $10^{3}\times h/M$ \\
\hline
\linf & 128, 192, 256      & $19.7$, $13.1$, $9.8$  \\ 
\ltwo & 128, 192, 256, 384 & $19.7$, $13.1$, $9.8$, $6.6$  \\
\fd   & 128, 192, 256      & $19.7$, $13.1$, $9.8$ \\
\hline
\end{tabular}
\end{center}
\label{tab:amr_runs}
\end{table}

\begin{figure}
\centering
	\includegraphics[width=\linewidth,clip=true]{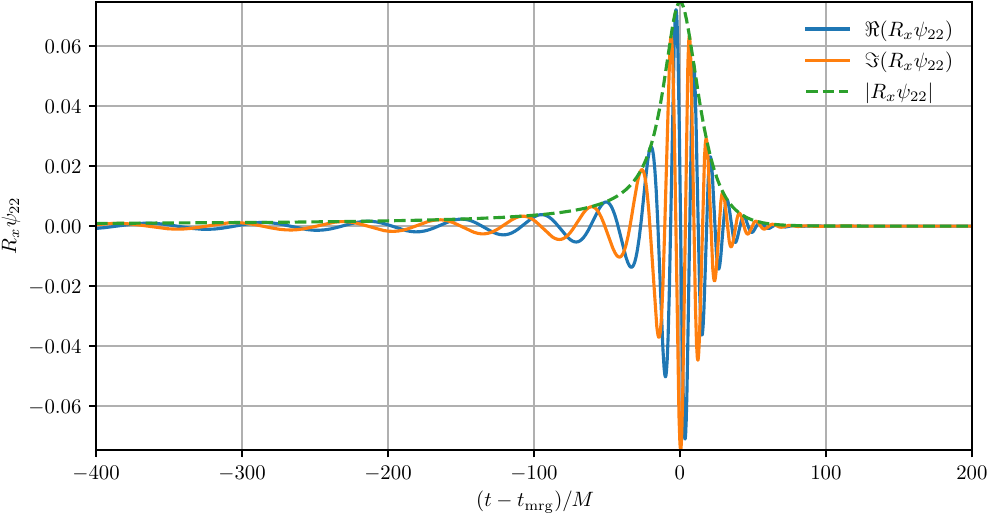}
	\caption{%
    Real~$\Re$ and imaginary~$\Im$ parts of the extracted $\psi_{22}$ 
    mode at $R_{x}=80M$ for the highest-resolution
    \ltwo{} simulation. The amplitude shown with the dashed line. We 
    use the maximum amplitude of the mode to determine the merger time 
    $t_{\rm mrg}$.
	}
\label{fig:psi22_ex}
\end{figure}

\begin{figure}
\centering
	\includegraphics[width=\linewidth,clip=true]{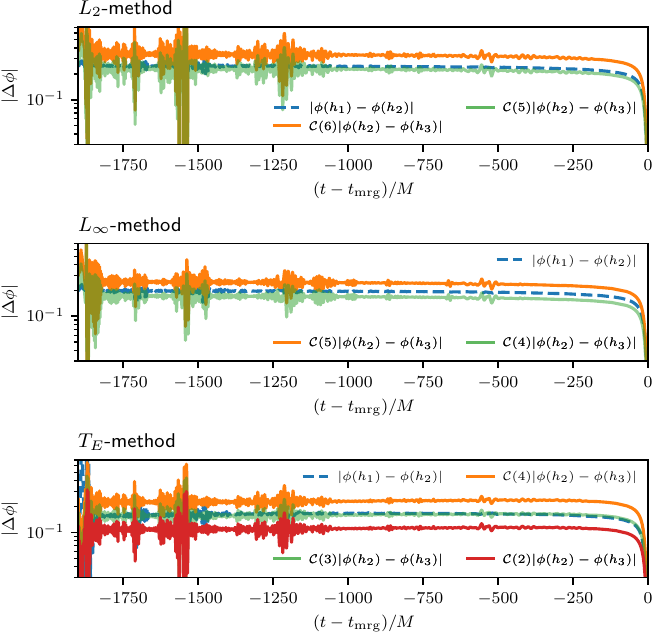}
	\caption{%
	Convergence tests of $\phi$ in $\psi_{22}(t) = A(t) e^{-i\phi(t)}$ for
	different \ac{AMR} criteria. Here $t_{\rm mrg}$ specifies the merger time.
    The resolutions $h_{1}, h_{2}, h_{3}$ are given in \Cref{tab:amr_runs}.
	We try different convergence order $n$ to 
	ascertain which line minimizes \Cref{eq:conv_order}.
	These lines illustrated by solid lines.
	The top panel shows the convergence of $\phi$ when \ltwo~method is used. 
	This method results in a fifth-order convergence.
	The middle panel exhibits a fourth-order convergence of \linf~method.
  The bottom panel depicts a third-order convergence when the \fd~criterion
	chosen for the simulation.
}
\label{fig:conv_test_phi}
\end{figure}

We employ the Newman-Penrose scalar
$\Psi_4$~\cite{Newman:1961qr} to extract \acp{GW} from our simulations, where
$\Psi_4 = \left(\ddot{h}_+ - i \ddot{h}_\times\right)$
and $\ddot{h}_+$ and $\ddot{h}_\times$ respectively denote
the second time derivative of the \ac{GW} strains: $h_+(t)$ and $h_\times(t)$.
In particular, we extract $\psi_{lm}$ modes at the radius $R_{x}$ to
construct $\Psi_4$ following~\cite{Bruegmann:2006ulg}
\be
	\Psi_4 = 
	\sum_{\ell=2}^{\infty}\sum_{m=-\ell}^\ell \psi_{\ell m}(t)\;
  {}_{-2}Y_{\ell m}(\vartheta,\varphi),
  \label{eq:psi_lm}
\ee
in which ${}_{-2}Y_{\ell m}(\vartheta,\varphi)$ stands for the spin 
weighted spherical harmonic bases. We show $\psi_{22}$ extracted at 
$R_{x}=80$ for the highest-resolution \ltwo{} simulation in \Cref{fig:psi22_ex}.
Additionally, one can write the dominant $\psi_{22}$ mode
\be
\psi_{22}(t) = A(t) e^{-i\phi(t)},
\ee
where $\phi$ denotes the phase of the waveform and $A$ its amplitude.
We obtain the error estimates of numerically calculated $\psi_{22}$
by assuming the error of the numerically calculated quantity of
interest $q$, can be modeled in powers of $h$, namely,
\be
  q(h) = q_{\rm ex} + c_n h^n + \mathcal{O}\big(h^{n+1}\big),
  \label{eq:q_expand}
\ee
where, $q_{\rm ex}$ is the exact value of the quantity, $n$ is the convergence
order, and $c_n$ is the constant coefficient of the leading-order error
term. Therefore, ignoring higher order terms of the error, 
the order of convergence $n$ can be estimated by minimizing the residual of
\be{}
\big| q(h_{1}) - q(h_{2}) \big| - \mathcal{C}(n) \big| q(h_{2}) - q(h_{3}) \big|,
 \label{eq:conv_order}
\ee
where $\mathcal{C}(n)$ is defined
\be
  \mathcal{C}(n) =
  \frac{\big| q(h_{1}) - q(h_{2}) \big| }
       {\big| q(h_{2}) - q(h_{3}) \big| }
   =
	\frac{ h_{1}^{n} - h_{2}^{n} }{ h_{2}^{n} - h_{3}^{n} },
 \label{eq:conv_order_c}
\ee
in which, subscripts 1, 2, and 3 refer to different resolutions. 
Accordingly, after aligning the phases of different resolutions at the merger
time $t_{\rm mrg}$,
we plot $\big| q(h_{1}) - q(h_{2}) \big|$
and compare it with different values of $n$ in
$\mathcal{C}(n) \big| q(h_{2}) - q(h_{3}) \big|$
in order to estimate what value of $n$  minimizes \Cref{eq:conv_order}.
Fig.~\ref{fig:conv_test_phi} shows the result of such convergence tests 
for different \ac{AMR} criteria where it depicts that 
$n \approx 5$, $n \approx 4$, and $n \approx 3$ minimize \Cref{eq:conv_order} 
for the \ltwo~method, \linf~method, and \fd~method respectively.

In our analysis, we aim to evaluate the impact of various sources of errors
within our numerical schemes, taking into account different strategies for
\ac{AMR}, and to determine which \ac{AMR} strategy
delivers faster convergence, allowing us to efficiently obtain accurate~\ac{GW}.
We further ask whether the observed
convergence order in \Cref{fig:conv_test_phi} aligns with theoretical
expectations.

Our numerical scheme introduces four sources of errors.
The first source pertains to the truncation error of the finite difference
operator, as elaborated in \Cref{sec:amr_fd}. This error is of order
$\mathcal{O}(h^{6})$.
The second source is associated with the Kreiss-Oliger dissipation
operator~\cite{alcubierre_book}. In our simulations, this operator employs
four ghost zones, resulting in an error of $\mathcal{O}(h^{7})$ ~\cite{Daszuta:2021ecf}.
Furthermore, our time integration employs the fourth-order Runge-Kutta
method. Consequently, the accumulated error, particularly noticeable during
extended time evolutions, scales with $\mathcal{O}(h^{4})$.
The fourth source of error arises from the prolongation operator
and amounts to $\mathcal{O}(h^{6})$. It is worth noting that in our
vertex-centered grid, the restriction operator is exact, as discussed
in \Cref{sec:prolongationrestriction}. As such the convergence order
should theoretically fall between $\mathcal{O}(h^{6})$ and $\mathcal{O}(h^{4})$.

When examining the data presented in \Cref{fig:conv_test_phi}, top 
panel, we can discern an approximate fifth-order convergence for the 
phase within the \ltwo~method.
Conversely, 
\Cref{fig:conv_test_phi},  middle panel, suggests a fourth-order 
convergence for the \linf~method. We speculate that this method 
exhibits a lower convergence order compared to the \ltwo~method due to 
the increased number of calls to the prolongation operator routines. As 
illustrated earlier in \Cref{fig:mbs_vs_time}, the \linf~method, when 
compared to the \ltwo~method, consistently generates and destroys 
\acp{MB} at a rate more than 400 times larger than the \ltwo~method at 
each time instance. Even though the prolongation operator is sixth 
order accurate, the frequent regridding reduces the smoothness of the 
numerical solution in time, and might amplify the $\mathcal{O}(h^{4})$ 
error term from the time integrator. Finally, the \fd~method, 
\Cref{fig:conv_test_phi}, bottom panel, shows third-order convergence, 
which may appear lower than the theoretical values. However, as 
explained in \Cref{sec:amr_fd}, the \fd{} method is tuned to have 
truncation error $\mathcal{O}(h^{4})$. The accumulation of this error over 
$\mathcal{O}(1/h)$ time steps is responsible for the overall
third convergence order.

\subsection{Waveform accuracy}
\label{sec:waccuracy}

In this section we study the quality of the extracted \acp{GW} and 
their computational costs for each \ac{AMR} method. Here, we only focus 
on the accuracy of the waveform extracted at finite extraction radius, 
since this is what we can control with the \ac{AMR}. For \ac{GW} 
astronomy application other sources of errors, in particular 
finite-extraction error, need to be taken into account. Previously, in 
\Cref{sec:efficiency}, we note that each refinement strategy develops a 
different grid structure. Equivalently, different \ac{AMR} strategies 
result in distinct number of \acp{MB} for the same resolution. 
Therefore, instead of using the resolution, we use the total number of 
evolved grid points across all refinement levels $\mathcal{N}$ to 
define a more representative computational cost for our runs. 
Furthermore, to define the word ``quality", we use the mismatch measure 
$\mathcal M$ between two different $\psi_{22}$ modes from two different 
runs, named $\psi(f)$ and $\psi_{\rm ref}(f)$ in their frequency space 
representations. We use the mismatch $\mathcal M$ 
definition~\cite{Vaishnav:2007nm,Lindblom:2008cm}
\be
\mathcal M\left[\psi(f),\psi_{\rm ref}(f)\right] = 1 - \mathcal
O\left[\psi(f),\psi_{\rm ref}(f)\right],
\label{eq:mismatch}
\ee
where, $\mathcal O\left[\psi(f),\psi_{\rm ref}(f)\right]$ is the overlap between
the waves and defined
\be
\mathcal O\left[\psi(f), \psi_{\rm ref}(f)\right] = 
\frac{\langle \psi(f)|\psi_{\rm ref}(f) \rangle}
			    {\sqrt{\langle \psi(f)|\psi(f)\rangle
					       \langle
					       \psi_{\rm ref}(f)|\psi_{\rm ref}(f)\rangle}},
\label{eq:overlap}
\ee
in which the inner product $\langle\cdot|\cdot\rangle$ is
\be
\langle \psi(f)|\psi_{\rm ref}(f) \rangle = \int_{-\infty}^{\infty}
	{(\psi^\dagger(f')
	\psi_{\rm ref}(f')+\psi(f')\,\psi_{\rm ref}^\dagger(f'))}\,df'.
\label{eq:inner_product}
\ee
In \Cref{eq:inner_product}, The $\dagger$ sign denotes the complex 
conjugate operator, and $\psi(f)$ is the frequency space representation 
of the $(\ell=2,m=2)$ mode of $\Psi_{4}$ (Eq. \ref{eq:psi_lm}), 
obtained using the continuous Fourier transform of the time domain 
$\psi_{22}$ mode. In practice, we utilize a discrete Fourier 
transformation of the numerically derived $\psi$ mode which is real and 
sampled over a discrete time line; hence, the integral limits are set 
by the lowest and highest frequency of the transformation.

\begin{figure}
	\centering
	\begin{subfigure}{.49\textwidth}
		\centering
		\includegraphics[width=\linewidth]{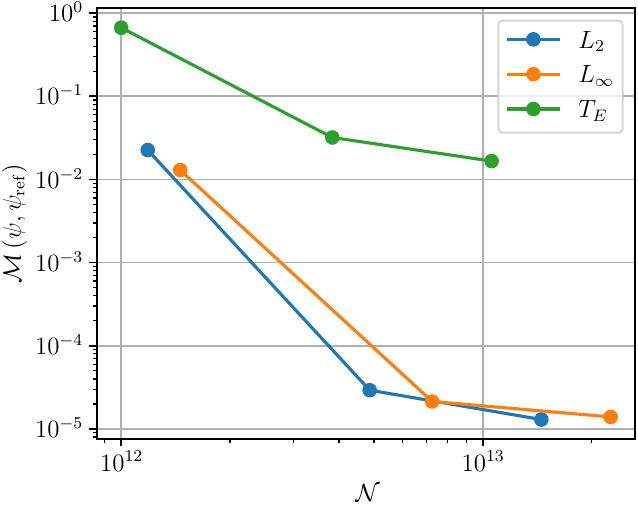}
	\end{subfigure}%
    \hfill
	\begin{subfigure}{.49\textwidth}
		\centering
		\includegraphics[width=\linewidth]{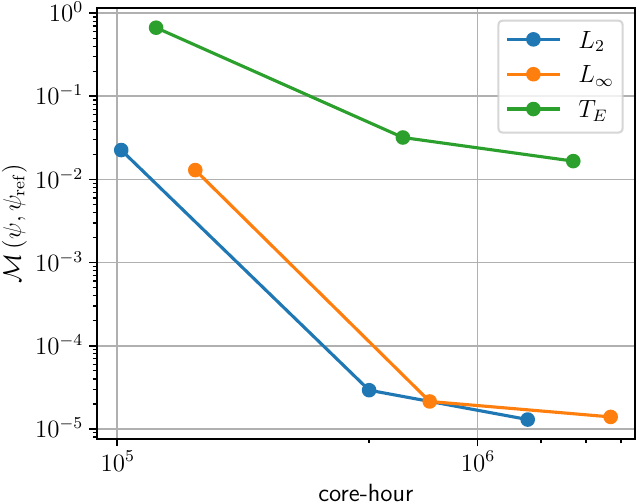}
	\end{subfigure}%
  \caption{\emph{Left panel:} mismatch \Cref{eq:mismatch} versus 
	the total number of grid point \added{updates} for different \ac{AMR} strategies.
	For the mismatch the benchmark waveform $\psi_{\rm ref}$ is computed using the \gra~mesh
  resolution $384$ with the \ltwo~strategy. For a given amount of the
  work, the lowest mismatch is yielded by \ltwo, 
	then by \linf, and finally by the \fd~method.
	\emph{Right panel:} mismatch versus the total core-hours
	taken by each \ac{AMR} criterion. For a given mismatch, the \ltwo~method takes
  less time and hence exhibits better performances with respect to the other
	\ac{AMR} methods.}
  \label{fig:mismatch_vs_work_and_walltime}
\end{figure}

To measure the accuracy and efficiency of our \ac{AMR} criteria in 
\gra, we carry out a high mesh resolution ($h_{4}=6.6\times 10^{-3}\, 
M$) simulation with the \ltwo~method,
\added{since this method has a higher convergence order
with respect to the other \ac{AMR} criteria.}
This simulation has $\mathcal N 
\approx 7\times10^{13}$. We use the resulting data $\psi_{\rm ref}$ as 
the benchmark for the mismatch comparison. We evaluate the mismatch 
\Cref{eq:mismatch} between the extracted $\psi$ from the simulations in 
\Cref{tab:amr_runs}~(the first three resolutions) and the $\psi_{\rm 
ref}$ wave. Fig.~\ref{fig:mismatch_vs_work_and_walltime} shows the 
result of this comparison. Each line corresponds to a different 
\ac{AMR} strategy and each point shows the resolution $\mathcal N$ and 
mismatch for each simulation. 

As can be seen from the left panel of \Cref{fig:mismatch_vs_work_and_walltime}, the 
\ltwo-criterion curve lies below both the \linf~and \added{\fd~curves}, 
meaning that it has a smaller error.
Even though the left panel of \Cref{fig:mismatch_vs_work_and_walltime} depicts the superiority of
\ltwo~method in terms of the total number of grid points $\mathcal N$,
it is not necessarily informative of the overhead taken
for refining and coarsening of the grid, i.e., interpolation and load
balancing. To put this into a perspective we also plot,
in the right panel of \Cref{fig:mismatch_vs_work_and_walltime},
the mismatch against the cpu core-hour
to determine how costly is each \ac{AMR} method.
As previously shown in \Cref{fig:mbs_vs_time}, \linf~has a larger number 
of \acp{MB} and a wider amplitude for creating and destruction of the \acp{MB}, 
so we expect \linf~to be more expensive
($\sim 60\%$ more expensive) than \ltwo~and \Cref{fig:mismatch_vs_work_and_walltime},
the right panel confirms this.
Notably, \Cref{fig:mismatch_vs_work_and_walltime} further shows \fd~is also
more costly than \ltwo~(despite the fact that is has lower number of
\acp{MB}). We speculate this extra cost comes from the extra
creation and destruction of \acp{MB}, \Cref{fig:mbs_vs_time}, 
with respect to the \ltwo~method and hence more calls to interpolation 
and load balancing routines.

Finally, we evaluate the potential of these methods to 
deliver sufficiently accurate waveforms for
the next-generation detectors such as LISA~\cite{LISA:2017pwj},
the Einstein Telescope~\cite{Maggiore:2019uih}, and
Cosmic Explorer~\cite{Reitze:2019iox}. 
To avoid biases in the parameter estimation for
\ac{GW} observations, we need to ensure the mismatch (from \Cref{eq:mismatch})
between the computed waveforms $\psi$ and the modeled one $\psi_{\rm ref}$,
satisfies  $\mathcal{M}[\psi,\psi_{\rm ref}] \leq p /(2 \rho^2)$,
where $p$ is the number of parameters used in the \ac{GW} model and $\rho$
is the signal-to-noise ratio of the detector~\cite{Chatziioannou:2017tdw}.
For next-generation
detectors signal to noise ratios as high as 1000 are possible 
and hence, the required mismatch is
$\mathcal{M} \lesssim 10^{-7}$.

Following Refs.~\cite{Ferguson:2020xnm, Vaishnav:2007nm}, we model the mismatch between $\phi$ at resolution $h$ and $\phi_{\rm ref}$ as
\be
\mathcal{M}\left(c,p,h,h_{\rm ref}\right) = 
c \left ({h}^p-h_{\rm ref}^{p}\right)^2,
\label{eq:mismatch_model}
\ee
where $c$ and $p$ are coefficients to be fit using the data. Using this 
ansatz, we can then estimate the mismatch between a waveform at given 
resolution $h$ and the exact solution as $\mathcal{M}(c,p,h,h_{\rm 
ref}=0)$.

\begin{figure}[h]
    \centering
    \includegraphics[width=\linewidth,clip=true]{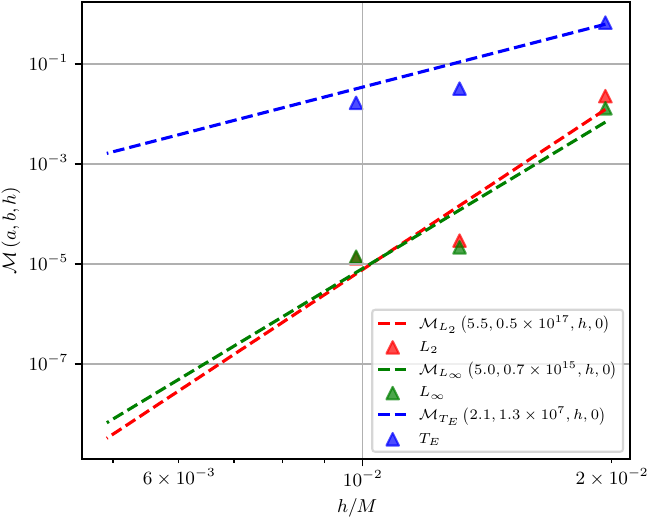}
    \caption{Mismatch models for different \ac{AMR} criteria. Here the data
    points shown with triangles with distinct color for each \ac{AMR}
    methods and 
		the dashed lines estimate the limit of the \Cref{eq:mismatch_model} at 
		$h_{\rm ref} \to 0$, i.e., infinite resolution for the benchmark run.
		The fastest convergence of mismatch obtained by \ltwo~method~(red color) 
		followed by \linf~(green) and then \fd~method~(blue).
	  }
    \label{fig:mismatch_xg}
\end{figure}

Fig.~\ref{fig:mismatch_xg} illustrates the result of this fit. We first
derive the fit parameters $c$ and $p$ using the 
data points~(shown with triangles in \Cref{fig:mismatch_xg}) obtained from 
the simulations, listed in \Cref{tab:amr_runs}.
The \ltwo~method yields the largest power ($p \approx 5.5$), followed 
by the \linf~method with $p\approx 5$, and finally \fd~method with 
$p\approx 2.1$. To further estimate the resolution required for a low 
mismatch, we assume the benchmark run has infinite resolution and 
hence, we take the limit of the fit for $h_{\rm ref} \to 0$. These are 
shown with dash lines in \Cref{fig:mismatch_xg}. While at low 
resolution the \linf{} method shows the smallest mismatch, we expect 
that the \ltwo{} strategy will result in the smaller errors at high 
resolution, thanks to its higher convergence order $p$.

Finally, we predict the minimum mesh resolution that is required for 
each \ac{AMR} criterion to yield a mismatch $\lesssim 10^{-7}$ as follows:
for the \ltwo~method $h/M = 6.57 \times 10^{-3}$,
in the \linf~method $h/M = 6.31 \times 10^{-3}$,
and in the \fd~method an impracticable resolution~($h/M = 0.51 \times 
10^{-3}$). It is notable that the model predicts that our highest 
resolution of \ltwo~method already achieves the desired low mismatch. 
However, we remark that this estimate does not account for 
all sources of error, but only for finite-differencing error,
\added{and the accuracy of this estimate depends on a limited amount of data
points.}
As such we favor the \ltwo~method as an efficient criterion to be used for the 
next-generation detectors.

\section{Discussion}
\label{sec:discussion}

We have presented a comparison of three \ac{AMR} strategies for 
\ac{BBH} merger simulations in the puncture code \gra. Two of the 
approaches set the grid resolution based on the distance from either of 
the two punctures. The first uses the maximum norm to determine the 
distance, \linf{}, and produces grid structures similar to those 
adopted by patch based \ac{AMR} codes, such as those based on the 
\texttt{Carpet} mesh refinement infrastructure \cite{Schnetter:2003rb} 
in the Einstein Toolkit \cite{EinsteinToolkit:2023_05}. The second 
approach uses the \ltwo-norm which better conforms with the geometry of 
the fields in a \ac{BBH} simulation. The third approach, here denoted 
as \fd{}, is based on an estimate of the local truncation error of the 
scheme. This method is, in principle, agnostic to the location of the 
puncture. In practice, we had to impose additional constraints on the 
grid structure that are informed by the position of the puncture, as 
discussed in Sec.~\ref{sec:efficiency}. While our results are specific 
to our code, we expect that the lessons learned will be applicable to 
other numerical relativity codes using similar techniques 
(finite-differencing, puncture methods).

We have performed simulations of an equal mass non-spinning \ac{BBH} 
system extending for ${\sim 20}$ \ac{GW} cycles up to merger and through 
merger and ringdown. We find that, for a fixed resolution on the root
grid, switching from the \linf{} to the \ltwo{} approach results in a 
${\sim} 60\%$ decrease in the number of mesh blocks (and hence grid 
points). The truncation-error based \ac{AMR} further reduced the number 
of grid points by ${\sim}30\%$. However, despite the significantly 
reduced number of grid points, the truncation-error based \ac{AMR} 
simulations are found to be more expensive in terms of actual number of 
core-hours than the \ltwo{} simulations. This is because \ac{AMR} results 
in frequent regridding and load balancing operations which slow down 
our simulations. The ``traditional'' \linf{} approach is the worst 
performer. It generates the largest number of mesh blocks and also 
incurs more frequent regridding than the truncation-error based 
\ac{AMR} simulations.

We have performed simulations at at least three different resolutions, 
spanning a factor of two in the grid spacing, for each \ac{AMR} 
strategy to verify convergence of our simulations. We find relatively 
clean fifth order convergence for the \ltwo{} simulations. The 
convergence order drops to fourth and third for the \linf{} and \fd{} 
\ac{AMR} strategies, respectively. The drop in convergence order for 
the \fd{} scheme should have been expected, because we set the parameters 
of the scheme so that the local truncation error of the scheme is 
$\mathcal{O}(h^{4})$. However, due to error accumulation over 
$\mathcal{O}(1/h)$ time steps, the convergence order is only 
$\mathcal{O}(h^{3})$. The reason for the drop in convergence order for 
the \linf{} scheme is unclear. We speculate that it might arise from 
the error accumulation in the time-integration scheme arising from the 
more frequent regridding.

The difference in convergence order is reflected in a better accuracy 
for the waveforms computed using the \ltwo{} \ac{AMR} strategy. To 
quantify this, we compute the mismatch between each simulation and a 
reference very high-resolution calculation. Both \linf{} and \ltwo{} 
simulations achieve mismatches $\mathcal{M} \sim 10^{-5}$ with the 
reference simulation with $h = 9.8\times 10^{-3}\, M$ on the finest 
grid. However, at this fixed resolution on the finest grid, and roughly 
at the same accuracy, the \linf{} simulations is $\sim 60\%$ more 
expensive than the \ltwo{} simulations. The truncation-error based AMR 
simulations have mismatch with the reference simulation larger than 
$10^{-2}$ are are found to be not competitive.

We cannot exclude that with additional tweaking to the refinement 
criteria, the \fd{} \ac{AMR} strategy could become competitive, 
e.g.,~Ref.~\cite{Radia:2021smk}. However, in our test the \ltwo{} 
\ac{AMR} strategy clearly outperformed more traditional ``box-in-box'' 
strategies both in terms of costs and accuracy. 
\added{Moreover, while we have only considered \linf~and \ltwo~in
this study, from a geometric standpoint, we anticipate to observe 
similar box-in-box or sphere-in-sphere characteristics when employing 
a general p-norm.
Nevertheless, using other norms may lead to varying levels of performance and
accuracy and hence it is a future research extension.}
Our results suggests 
that \gra{} can produce waveforms with mismatches as small as 
$10^{-7}$, sufficiently accurate for next-generation \ac{GW} 
experiments, using the \ltwo{} AMR strategy. An important future 
extension for this work is to design \ac{AMR} strategies for high 
mass-ratio \ac{BBH} simulations.

\ack

This work was supported by NASA under award No. 80NSSC21K1720 and the 
U.S. Department of Energy, Office of Science, Division of Nuclear 
Physics under Award Number(s) DE-SC0021177.
Simulations were performed on Bridges2, Expanse (NSF XSEDE allocation 
TG-PHY160025), Frontera (NSF LRAC allocation PHY23001), Perlmutter, and 
on the national HPE Apollo Hawk at the High Performance Computing 
Center Stuttgart (HLRS). This research used resources of the National 
Energy Research Scientific Computing Center, a DOE Office of Science 
User Facility supported by the Office of Science of the U.S.~Department 
of Energy under Contract No.~DE-AC02-05CH11231.


\bigskip
\noindent
{\bf References\\}

\bibliographystyle{unsrt}

\end{document}